# Into Summarization Techniques for IoT Data Discovery Routing

Hieu Tran, Son Nguyen, I-Ling Yen, Farokh Bastani
*Department of Computer Science*
*The University of Texas at Dallas*
Richardson, Texas, USA
{trunghieu.tran, sonnguyen, ilyen, Farokh.Bastani}@utdallas.edu

*Abstract*—In this paper, we consider the IoT data discovery problem in very large and growing scale networks. Specifically, we investigate in depth the routing table summarization techniques to support effective and space-efficient IoT data discovery routing. Novel summarization algorithms, including alphabetical based, hash based, and meaning based summarization and their corresponding coding schemes are proposed. The issue of potentially misleading routing due to summarization is also investigated. Subsequently, we analyze the strategy of when to summarize in order to balance the tradeoff between the routing table compression rate and the chance of causing misleading routing. For experimental study, we have collected 100K IoT data streams from various IoT databases as the input dataset. Experimental results show that our summarization solution can reduce the routing table size by 20 to 30 folds with 2-5% increase in latency when compared with similar peer-to-peer discovery routing algorithms without summarization. Also, our approach outperforms DHT based approaches by 2 to 6 folds in terms of latency and traffic.

## I. INTRODUCTION

The rapidly increasing number of IoT devices results in the creation of a torrent of data. These data may be stored and managed at the edge or regional sites of the Internet. For example, many smart cities may host their IoT data in the municipal databases. Many manufacturers may host their IoT data at their own sites. Data from smart vehicles will probably stay with the vehicles or at the roadside units. As can be seen, a lot of IoT data are being created and hosted in a dispersed way over the Internet. These peer-to-peer (p2p) data sources can be discovered and made use of to achieve more advanced data analysis and knowledge discovery.

To enable IoT data stream discovery, annotation is an important issue. Discovery literature has considered raw keywords and multi-attribute annotation (MAA) models for annotation [1]. Many machine-interpretable semantic models for annotating the IoT data streams, such as the semantic sensor network, Fiesta-IoT, etc., have been proposed. However, complex annotations may significantly degrade the IoT data discovery performance. In this work, we consider MAA based IoT data annotation and query specifications. The semantic based annotations can be converted into MAA to facilitate efficient routing.

Data discovery in p2p networks under keywords and MAA based annotations has been widely investigated. Many existing works are DHT (distributed hash table) based [2], which hashes data objects to specific nodes in the network. They are very effective for some objects but are not suitable for IoT data. Basic DHT schemes require data to be stored at the hashed locations. IoT data streams are generally continuous flows and moving them to potentially far away hosts based on DHT can cause major overhead. Some works use the hashed nodes as the pointers pointing to the actual IoT data locations [1] [3], but these approaches imply space and communication overhead, especially in handling MAA based lookups.

Unstructured discovery routing approaches, such as [4] [5], require the maintenance of a routing table or routing information cache. These discovery routing schemes are similar to IP-based routing, except that the routing tables and the discovery queries are indexed by keywords instead of IP addresses. One important issue in unstructured p2p routing is the design of the routing table for maintaining or caching the routing information. Since many IoT data streams are hosted by resource-constrained nodes at the edge of the network, routing table size on these nodes has to be confined. In IP networking, space concern is addressed by compressing the routing tables using the "summarization" technique [6]. However, very few works have considered space efficient routing table design or summarization techniques.

In this paper, we consider data discovery in p2p IoT networks at the edge of the Internet (including resource-constrained edge servers and IoT nodes). We focus on the space efficiency issue in query routing for discovery of data streams annotated by the MAA model. Specifically, we use the summarization techniques to achieve space efficiency, which have been widely used in IP-based routing.

**How to summarize keywords used in data discovery routing**? Summarization in IP-based routing relies on the hierarchical structure of the domain names [6]. Summarization of numerical values can also be done intuitively by union of individual ranges [5]. For summarizing keywords, a natural extension of these methods is *alphabetical based* summarization, which compresses multiple keywords into their longest common prefix. However, alphabetical based method depends on how likely we can find proper keyword groups that can be summarized into common prefixes. If the group of keywords that can be summarized does not show up together in a routing table, then it will be difficult to compress the table. In fact, experimental results show that the effectiveness of alphabetical based summarization is limited.

To avoid the dependency on keyword distributions, we can use hashing to randomize the set of keywords in the system to a uniformly distributed coding space to allow summarization to occur in a more uniform manner. Accordingly, we also consider the *hash based* summarization in this paper.

When considering keyword distributions carefully, we can see that relevant keywords may have some regionalized appearance. For example, some cities are major manufacturing sites for certain products, such as Detroit for automotive manufacturing, Chicago for steel industry, etc. Specific sensor data streams for production line monitoring for the specific industries may include many semantically similar keywords. In a smart city setting, sensors with similar functionalities, such as for traffic observation, pollution monitoring, trash tracking, etc., may be used throughout the city. The descriptors for these sensor data streams will also have significant similarities. To achieve potentially better routing table compression by making use of the regionalized keyword similarity distributions, we also consider the *meaning based* summarization to summarize IoT data descriptors with similar keywords and expect that it will yield a better compression ratio.

**Is summarization for keywords feasible**? With keywords indexing the routing tables, how do we know which keywords (routing table entries) can be aggregated? This is straightforward for IP based routing, for numerical ranges, and for the alphabetical based method, but difficult in other approaches. Having keywords such as temperature, pressure, flow rate, etc., in a routing table, how do we know whether they can be summarized? No work has specifically addressed this issue.

We design novel coding schemes to maintain parent-child summarization relations to enable the identification of the child keywords that can be summarized and the parent keyword to be summarized into. These coding schemes can further improve space efficiency of the routing tables. Thus, we allow routing information to be summarized instead of thrown away to achieve space efficiency as well as routing effectiveness.

**How to know when to summarize**? Unlike IP based routing, where same domain prefixes have a high probability of implying the same routing directions, keyword based summarization does not have the same implication. Assume that a parent keyword $p$ has three child keywords $x$, $y$, and $z$ (i.e., $x$, $y$, and $z$ can be summarized into $p$). During summarization, how do we know $p$ has three child keywords? This is an issue even for alphabetical based summarization. Again, we need to maintain the ontology at each routing node, which is not feasible. Also, should we require that summarization to $p$ can only be done when all the 3 child keywords appear in the routing table (full coverage) or allow a partial set of these keywords to be summarized into $p$ (partial coverage). With IP addressing, this is less a problem due to its domain based design. But in keyword based routing, requiring full coverage reduces the routing table compression ratio and allowing partial coverage may result in misleading routing. In this paper, we not only analyze the impact of the "coverage" threshold, but also develop novel schemes to compute coverage for truly realizing the summarization strategies.

**Experimental study**. We crawled the web, obtained 100K IoT data streams, and extracted their descriptors [7]. Accordingly, we conducted in-depth experiments to study the performance of our summarization strategies under different settings. We also compared our approach with DHT-based and other unstructured routing solutions in large-scale IoT networks. Experimental results show that our summarization solutions yield great space efficiency for routing tables.

In the rest of the paper, Section II discusses the related literature. In Section III, we formally define our multi-attribute annotation model and our discovery routing problem. Section IV discusses the three summarization strategies and the algorithmic designs to realize them. Section V considers the summarization coverage issue and discusses the schemes to enable coverage computation. Experimental results are presented in Section VI. Section VII concludes the paper.

## II. RELATED WORKS

Centralized IoT data discovery solutions [8], [9] require tremendous amount of resources to maintain the information of all the IoT data sources all over the world. With the fast growing number of IoT devices and their corresponding raw and processed data, the scalability may become a concern. More importantly, if the change rate of the IoT data streams are significant, it will take a lot higher communication cost to keep the centralized information up to date. Also, locality sometimes is important in the discovery process. Thus, peer-to-peer (p2p) discovery still plays an important role. There are two major directions in p2p data discovery, DHT based and unstructured routing, which were mainly designed for document search.

**DHT based solutions for MAA based data discovery**. A lot of works in keyword based discovery in peer-to-peer systems are based on distributed hash table (DHT) [2] [10]. However, DHT, in principle, only handles discovery based on a single keyword. There are several works that extend DHT for handling object lookup with each object indexed by multiple keywords or MAA. Some works consider hashing individual keywords to the DHT. The querier has to retrieve all the matching object indices for each keyword in the query, determine the real match from the responses, and retrieve the object. [3] is such a solution. This approach incurs a heavy communication cost, i.e., $l$ round trips of communication for a query with $l$ keywords. Also, a lot of information passed back may not be useful toward the query responses.

Some other DHT-based works consider hashing all the powersets of the keyword set (or attribute-value pairs) for each object into the DHT. For example, each object may be specified by 10 attributed keywords, but the query may give any subset of the 10 keywords. Thus, $2^{10} - 1$ keyword sets are hashed to the DHT to serve as pointers. [1] [11] are solutions based on this approach with various improvements.

[12] introduces the MKey scheme, which is a two-level approach. At the high level, a set of super-peers are structured by DHT. Each super-peer is associated to a cluster of peers and unstructured routing is used in the cluster. Same as other multi-keyword DHT schemes, object indices are still duplicated, and lookup needs to explore some DHT clusters that may not have the matching object.

Though DHT based approaches maintain exact information of where to forward a query, most of them still require $\log N$ overlay routing steps and each overlay step traverses half of the network ($= 1/2\sqrt[k]{N}$ nodes, where $k$ is the connectivity degree) in average. For document search, DHT requires the objects being placed at their hashed locations, which is infeasible for IoT data discovery since IoT data streams are collected continuously. Moreover, basic DHT only handles a single keyword and MAA based approaches, as discussed above, further introduce space and communication overheads.

**Unstructured peer-to-peer networks and summarization techniques**. Information centric networking (ICN) consists of a set of keyword-based unstructured discovery routing schemes. In DONA [13], a public key based naming has been considered. It uses centralized or hierarchical Resolution Handler (RH) for naming resolution and IP like forwarding. In Content Centric Networking (CCN) [14] [15] and its successor Named Data Networking (NDN) [16], the object naming is by the URI like hierarchical descriptors and the routing protocols are similar to the conventional IP based approaches. These naming schemes are not suitable for the discovery of IoT data streams. A hierarchical naming, though it can merge different attributes into one string, cannot handle missing attributes or allowing attributes being specified without a predetermined order. Also, none of the ICN works consider the feasible summarization techniques.

The GSD scheme [4] is an early unstructured peer-to-peer information caching solution and several similar works provide various improvements. None of these works consider summarization techniques or how to compress routing tables.

The Combined Broadcast and Content Based (CBCB) routing [5] is an unstructured peer-to-peer solution that can be used for object discovery. It considers a covering concept, i.e., if the newly advertised routing data is "covered" by existing routing data, then there will be no updates and the advertised information will not be forwarded further. Also, it merges multiple routing table indices if they can be expressed by an index covering all of them. Some claim this to be a form of summarization. But this intuitive technique can only be applied to numerical ranges effectively. When considering keywords, CBCB considers combining them by "and" and "or" operators, which may even increase the number of entries in the routing table. Thus, the space efficiency issue is not addressed.

Most of the unstructured p2p routing schemes adapt IP-based routing protocols by replacing IP addresses with keywords, but almost none of them consider the summarization techniques as their counterparts do because it is easier to summarize IP addresses, but difficult to summarize keywords. We use the AODV (ad hoc on-demand distance vector) IP-routing protocol with MAA but focusing on keyword-based summarization to reduce space overhead.

### III. EXAMPLES AND PROBLEM SPECIFICATION

#### A. *Example Cases*

We consider data discovery in the p2p IoT database network (IoT-DBN), which consists of a huge number of individual IoT data streams and databases (each hosts multiple data streams) dispersed over the regional and edge of the Internet. A few example cases are given to illustrate IoT-DBN and the system model.

**Case 1**. Consider a scenario where an ambulance is requested for some injured victim of an accident. The goal is to get the ambulance that can arrive at the site at the earliest time. Various Map Apps may estimate driving time for regular cars, but not for ambulances. If the data is not directly available, we may collect ambulance driving data from other cities in similar traffics to derive the potential ambulance driving time.

**Case 2**. Consider that a police officer receives a report of a hit-and-run incident that occurred 5 minutes ago by a red sedan at a location with coordinate $(x, y)$. The police officer can request the recorded videos, if available, from vehicles that are within 5-minute driving range from $(x, y)$ to track the suspect.

**Case 3**. Performing data analytics based on multiple similar data sources can help significantly improve the analysis accuracy and help with cold start situations. For example, a combustion engine production company may retrieve sanitized data from noncompeting companies with similar production lines to help with its own analytics for early fault detection and diagnosis of its production line. If the company is new and does not have its own data, existing data can help avoid cold start.

**Example for the IoT-DBN model**. Consider Case 1 with sample GPS data for real-time traffic tracking provided by, for instance, the City of Cincinnati, the City of Columbus, or the City of Cleveland, etc. The annotation for a sample data stream is shown below.

(DataCategory: *GPS*; Vehicle type: *car*;
 City: *Cincinnati*, Region: *Central Business District*;
 Day: *Weekend*; Region traffic volume: *v*;
 Duration: *3/2/18 17:03:20 - 3/2/18 19:10:30*)

The data streams may be hosted by individual cities and may still be in collection or may have finished collection. (Note: for Examples 2 and 3, the data are hosted by individual cars and individual companies, respectively.) Each IoT data stream is annotated, and the annotation (aka metadata) is a set of attribute-value pairs (descriptors). The example annotation contains static as well as dynamic attributes. The dynamic attributes are continuously updated as the new data flow in. "Region traffic volume" is an example of dynamic attributes, specified as the average number of vehicles per unit time passing through this region and computed up to the current time if the duration has not ended. Attribute "Vehicle type" is an example of static attributes and for Case 1, we need to consider types *car* and *ambulance*. (Note: Individual data objects in the data streams are annotated by GPS, timestamp, etc., but are not part of the data stream description.)

To enable IoT-DBN capabilities, we add a thin routing layer above individual distributed IoT data streams and/or databases to parse the data streams for attribute consistency and to construct routing tables and perform discovery routing.

Following Case 1, a policeman from, e.g., Indianapolis, may issue queries to retrieve GPS data for vehicles of type *sedan* and *ambulance* of nearby cities with any time period to derive the correlations and predict the driving time for ambulances in the accident region (which is assumed to be not available). The query could be described as following:

(DataCategory: $GPS$; Vehicle type: $ambulance \;||\; car$;
Region traffic volume: $[v_1, v_2]$)

The query requests data from any region and any duration with a traffic volume that is similar to the current condition. It may return Cincinnati's ambulance and car GPS data, which can help build the mapping between their driving times. The mapping can then be used to predict the ambulance driving time in Indianapolis assuming that only car GPS data is available in this city and the ambulance data is missing. We applied transfer learning techniques to make a similar prediction (for service vehicles instead of ambulance). It yields a high accuracy (MAPE=13.8) and takes only a few seconds to derive the mapping (based on 380K data points) and make the prediction. (The mapping is derived at the data source and only the model properties need to be transferred.)

### B. Problem Specification

We consider the data discovery routing problem in IoT-DBN. For a data stream $ds$, let

$$AN^{ds} = \left((a_1{:}v_1^{ds}), (a_2{:}v_2^{ds}), \ldots, (a_n{:}v_n^{ds})\right)$$

denote the annotation (key-value pairs) for $ds$, where $(a_i{:}v_i^{ds})$ is the "*descriptor*" for the $i$-th *attribute* $a_i$, and $v_i^{ds}$ is the *attribute value* for $a_i$ in $ds$'s description. Note that $AN^{ds}$ specifies a data stream $ds$ that plays a role as its *metadata* for discovery purpose, not individual data points. And, data discovery is based on the descriptions of data, i.e., $AN^{ds}$.

Let $AV_{a_i}^{sys}$, $1 \le i \le n$, denote the set of descriptors (attribute-value pairs) for attribute $a_i$ in $AN^{ds}$, for all $ds$ of all involved IoT databases in an IoT-DBN. $AV_{a_i}^{sys}$ may change dynamically, but it is relatively stable, and we consider $AV_{a_i}^{sys}$ at a specific time. We assume that an IoT-DBN is a specific application domain and there exists a common set of attributes. Even if some different attributes may appear, they are aligned, and the union of the attributes are considered. Also, it is possible that for some data streams, the values for some attributes are missing, i.e., $v_i^{ds} = \emptyset$.

Users may issue queries to look for data streams in the IoT-DBN. A query is specified by descriptors with a matching threshold. Specifically, $q = (A^q, \alpha, q_{src})$, where

$$A^q = \left((a_1{:}v_1^q), (a_2{:}v_2^q), \ldots, (a_n{:}v_n^q)\right),$$

$\alpha$ is a threshold for query matching (will be defined later) and $q_{src}$ is the querier node. Also, $v_i^q = \phi$ if $a_i$ is not given in $q$.

**Attr-match**. Consider a descriptor in a query $q$, $(a_i{:}v_i^q)$, and a descriptor for a data stream $ds$, $(a_i{:}v_i^{ds})$. They have an attr-match for attribute $a_i$ iff $v_i^q = v_i^{ds} \vee v_i^q = \phi \vee v_i^{ds} = \phi$. Note that we can define operation "=" differently for different usage requirements.

In IoT-DBN, each database node $D$ maintains a routing table $RT^D$. $RT^D$ consists of a set of entries $\left((a{:}v), D'\right)$, where $(a{:}v)$ is a descriptor and $D'$ is the forwarding neighbor through whom a node hosting a data stream $ds$ with matching $(a{:}v)$ can be reached. Note that the entries in $RT^D$ are organized in a certain data structure for efficient lookup (will be discussed in Sec. VI).

**α-match**. Consider a query $q$ and a data stream $ds$ ($q$ and $ds$ are as defined earlier). We define $match_i = 1$ if $(a_i{:}v_i^q)$ and $(a_i{:}v_i^{ds})$ have an attr-match, and $match_i = 0$, otherwise. $ds$ has an $\alpha$-match with $q$ if and only if $\sum_{i=1}^{n} match_i \ge \alpha \times n$, $0 \le \alpha \le 1$. When $\alpha = 1$, each attribute provided in $q$ should have a match in $ds$. Similarly, consider an IoT-DBN node $D$. If $\exists D'$, such that $RT^D$ includes at least $\alpha \times n$ different entries $\left((a_i{:}v_i^q), D'\right)$, for some $i$, then $D'$ is an $\alpha$-match forwarding neighbor of $D$ for $q$.

**Discovery routing**. Our goal is to perform discovery routing for query $q$ to discover data streams $ds$ by matching data stream annotations $AN^{ds}$. A routing mechanism consists of the advertisement protocol and the query forwarding protocol. We consider both table based routing and on-demand information caching (we call information cache also as routing table for consistency). Also, we consider the common routing protocol with our goals and assumptions.

*Advertisement protocol (AdvP)*. The advertisement phase is for routing table construction. Generally, it is performed when there are new data streams. During advertisement, node $D$ sends an advertisement message $advmsg$, which includes the descriptors of new data streams, to all its neighbors. When $D'$ receives $advmsg$ from $D$, for each descriptor $(a{:}v)$, an entry $\left((a{:}v), D\right)$ is added to $RT^{D'}$ if it does not already exist. The system gives a bound $b^{ad}$, which is the number of hops an advertisement message can traverse. If the hop count for $advmsg$ has not reached the bound $b^{ad}$, it will be forwarded to all $D'$'s neighbors except those listed as the forwarding neighbors for the descriptors in $advmsg$.

*Query Forwarding protocol (QFP)*. The system gives a bound on the hop counts for query forwarding, denoted by $b_q$. Consider a node $D$ receiving a query $q$ from $D'$. Let $R^{q,D} = (ds_1, ds_2, \ldots, ds_r)$ denote the data streams in $D$'s database that have $\alpha$-match with $q$. If $R^{q,D} \ne \phi$, then $D$ sends response $(R^{q,D}, q, D)$ back to $q_{src}$. In table driven routing, $R^{q,D}$ can be sent directly to $q_{src}$. In information caching based approaches, $R^{q,D}$ is delivered by a reverse route of query forwarding. Also, let $mns^{q,D} = \{D_1, \ldots, D_m\}$ denote the set of $\alpha$-matching neighbors for $q$ at node $D$. $D$ forwards $q$ to the neighbors in $mns^{q,D}$. If $\|mns^{q,D}\| = 0$ or the hop bound $b^q$ has been reached, then $D$ will not forward $q$. As can be seen, if there are no updates and $b^{ad} = \infty$ and $b_q = \infty$, then QFP will return all $\alpha$-matching data streams for $q$.

### IV. SUMMARIZATION TECHNIQUES

To enable data stream lookup in an IoT-DBN, a routing table $RT^D$ of node $D$ maintains $AN^{ds}$ for each data stream $ds$ whose advertisement reaches $D$. Since many nodes in the IoT-DBN may be resource-constrained, we consider space efficiency for routing tables. Specifically, we consider the summarization technique for routing table compression.

In general, attribute values of descriptors may be keywords or numerical values. Summarizing numerical values can be by union of numerical ranges like in [5]. We focus on summarization for *attributed keywords*. Also, for the MAA model, matchmaking for values of different attributes should be considered differently, so we should only summarize keywords of the same attribute. We consider different summarization policies: alphabetical based policy ($SP_{alph}$), hash based policy ($SP_{hash}$) and meaning based policy ($SP_{meaning}$). We use $SP_x$ to refer to either of these policies when discussing some strategies that are common for them.

The fundamental element in routing table summarization is which descriptors can be summarized and what they can be summarized into. To facilitate the definition of these summarization relations, we define the concept of the summarization tree. Let $ST_x(AV_{a_i}^{sys})$ denote the summarization tree (*sum-tree*) constructed from the set of all the descriptors in the IoT-DBN, $AV_{a_i}^{sys}$, for attribute $a_i$ using summarization policy $SP_x$ (sometimes abbreviated as $ST_x$). In $ST_x$, the child nodes (*st-children*) are candidates to be summarized into their parent node (*st-parent*) and summarization can be done recursively, from leaves to root.

$SP_{alph}$ is a straightforward extension of IP summarization. In $ST_{alph}$, an st-parent "string" is the longest common prefix (*LCP*) of its st-children "strings".

$SP_{hash}$ is designed to probabilistically improve the compression performance of $SP_{alph}$. Compression in $SP_{alph}$ depends on the probability of prefix match among a set of keywords in a routing table, which further depends on the keyword distributions in the application domain. We use hashing to map keywords to a uniform space. In $ST_{hash}$, an st-parent "hash code" is the LCP of its st-children "hash codes".

$SP_{meaning}$ considers the keyword distributions in the topology of IoT-DBN. IoT data streams collected by the sensors of a specific system or a specific environment may have attributed keywords that are semantically similar. If we summarize routing table entries based on the semantic meaning of the attributes, we may get better compression in some neighborhoods. For example, some cities are major manufacturing sites for certain products, such as Detroit for automotive manufacturing, Chicago region for iron and steel industry, etc. Specific sensor data streams for the production line monitoring for these specific industries may be generated. For instance, keywords like engine-speed, engine-pressure, air-fuel, crank-position, throttle-position, etc., are likely to appear in automotive manufacturing regions and they may be "summarized" due to their relevance and semantic similarities.

**Summarization trees**. The summarization tree maintains the parent-children relations of summarization, i.e., children nodes are to be summarized into the parent node. Also, instead of the raw attribute values (keywords), each node $t$ in $ST_x$ maintains a "*code*", denoted by $t.\tau_i$, which is derived from $t.v_i \in AV_{a_i}^{sys}$ or from the code(s) of its parent or children in $ST_x$. Let $SC_{a_i}(t^p) = \{t_j \mid 1 \leq j \leq t^p.nc\}$ denote the set of $t^p.nc$ children of a parent node $t^p$. Also, let $SCC_{a_i}(t^p) = $ $\{\tau_{i,j} \mid 1 \leq j \leq t^p.nc\}$ denote the set of codes maintained by st-children in $SC_{a_i}(t^p)$ and $\tau_i^p$ denote the code of $t^p$, i.e., $\tau_{i,j} = t_j.\tau_i$ and $\tau_i^p = t^p.\tau_i$. As can be seen, $\tau_{i,j}$, for all $j$, can be summarized into $\tau_i^p$.

$ST_{alph}$. In $ST_{alph}$ for attribute $a_i$, a leaf node with value $v_i$, $v_i \in AV_{a_i}^{sys}$, has code $\tau_i = v_i \cdot \$$, i.e., concatenating terminator '\$' to its value. Also, $\tau_i^p = LCP(\tau_{i,j}), 1 \leq j \leq nc$, where $LCP$ is the longest common prefix function. As can be seen, $ST_{alph}$ is a trie [17]. Note that without the terminator, we cannot differentiate a keyword from a summarization code. For example, consider a node with $\tau_i = $ CO and it has children $CO_2$ and $CO_3$. We will not know whether CO is a keyword in $AV_{a_i}^{sys}$ or just the code of an internal node in $ST_{alph}$.

$ST_{hash}$. In $ST_{hash}$, let the depth of the tree be $d$ and the maximal degree for any node be $2^c$. We have $(2^c)^d \geq \|AV_{a_i}^{sys}\|$. A leaf node $t$ of $ST_{hash}$ has code $t.\tau_i = h(t.v_i)$, where $h$ is a hash function mapping each attribute value into a code of $c \times d$ bits with a 1 added as the left-most-bit (a total of $c \times d + 1$ bits). Also, $\tau_i^p = \left\lfloor \frac{\tau_{i,j}}{2^c} \right\rfloor, 1 \leq j \leq nc$. The code of each st-parent is constructed by removing the last $c$ bits from the code of any of its st-children.

In $ST_{hash}$, $c$ and $d$ are configurable parameters. Larger $c$ and $d$ result in a sparser summarization tree, making compression less effective. On the other hand, smaller $c$ and $d$ results in a smaller hash space, limiting the extensibility of the coding scheme. Hashing could incur collisions, but it is equivalent to preliminary summarization.

$ST_{meaning}$. $ST_{meaning}$ is constructed by two operations, clustering and coding. First, we apply Word2Vec to map keywords in $AV_{a_i}^{sys}$ to the corresponding set of vectors, denoted as $w2v(AV_{a_i}^{sys})$. $w2v(AV_{a_i}^{sys})$ is considered as the *root* vector set, and it is coded by $root.\tau_i = 1$. Then, we cluster the root set into $2^c$ clusters of vector sets using the K-mean algorithm and each cluster becomes a child of the root. We recursively cluster each parent set with code $\tau_i^p$ into $2^c$ child clusters and the code of its $j$-th st-child will be $\tau_{i,j} = \tau_i^p \times 2^c + j$. Clearly, the resulting tree $ST_{meaning}$ keeps semantically similar keywords in the same subtree.

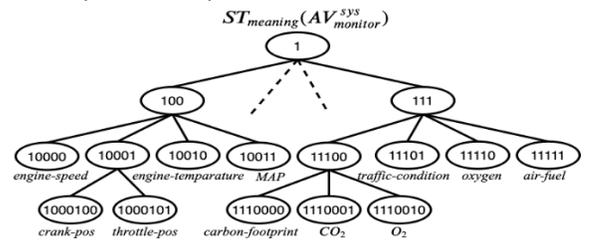

Fig. 1. Sample of $ST_{meaning}$

Fig. 1 shows the coding of an example $ST_{meaning}$ with $2^c = 4$. First, all keywords belong to the *root* node and $root.\tau_i$ is coded by 1. These keywords are clustered into 4 child nodes and their codes are generated by appending 00 to 11 to the parent code, i.e., $\tau_i = $100 to 111. The clustering process continues till a node has only $k$ keywords in its cluster and $k \leq 2^c$. If $k = 1$, then the node becomes a leaf node. If $1 < k \leq 2^c$, then the node generates $k$ children without clustering. Thus, each leaf node has only one keyword. Each child node appends its index to its parent code to obtain its $\tau_i$ value. Thus, nodes at levels 2, 3, and 4 are coded by 3, 5, and 7 bits, respectively.

**Make summarization for keywords feasible by summarization tree based coding**. If we maintain raw attribute values $(a_i, v_i)$ in routing tables, we will not be able to identify which child keywords can be summarized into which parent keyword unless we maintain $ST_x$ at each node in the network. But this will fully defeat the purpose of summarization. It will also make per node matchmaking very inefficient. The solution is to code $ST_x$ such that the parent-children summarization relations are carried by the code $(a_i, \tau_i)$. As discussed above, the coding design in our $ST_x$ can be used to index the routing tables to (1) enable summarization, (2) further reduce routing table size, and (3) support efficient matchmaking.

For $SP_{meaning}$ and $SP_{hash}$, a group of codes of length $len$ bits are siblings in $ST_x$ if they have a common prefix of length $len - c$. For example, in Fig. 1, codes 1000100 and 1000101 have the same prefix 10001 and they can be summarized into 10001. In $SP_{alph}$, the codes are keywords and, naturally, siblings are identified by their common prefix strings.

The code size constructed via $ST_{hash}$ and $ST_{meaning}$ are very compact. With 100K keywords and $2^c = 4$, we need a tree with $d = 9$ and code $t$. $\tau_i$ will be 19 bits.

**How can a node in IoT-DBN know the codes of its descriptors**? When new data streams are created at a node $x$ in IoT-DBN, all descriptors $(a_i : v_i)$ are transformed into $(a_i, \tau_i)$ and used in advertisement. When issuing a query, the querier also needs to map the attributed keywords in the query to the corresponding codes. For $SP_{alph}$ and $SP_{hash}$, $x$ can derive the code $\tau_i$ directly from the attribute value $v_i$ as discussed above. $x$ does need to know $c$ and $d$ values, which can be obtained from a neighbor when $x$ joins the IoT-DBN. For $SP_{meaning}$, we have to construct the tree $ST_{meaning}$ to derive the code for each attribute value. We cannot maintain $ST_{meaning}$ on each node, thus, we use a central summarization tree server $STS$ to construct and host $ST_x$. IoT-DBN peers can query $STS$ to get the code mapping for the descriptors it hosts. We can also replicate $ST_{meaning}$ on multiple super-peers to improve availability, distribute the load, and reduce communication cost for code mapping queries. Since code mapping queries are only done upon the creation of new data streams and discovery queries, the overhead will be reasonable.

**Handling growing IoT-DBN**. In IoT-DBN networks, new data points are generated continuously for each data stream, but the descriptors for each existing data stream do not change. However, new IoT devices may be introduced continuously which will generate new data streams. Also, new data analysis workflows will generate new data streams. Thus, $ST_x$ will grow which may result in potential changes in the code $\tau$. In $ST_{alph}$ and $ST_{hash}$, new code $\tau$ for the new data streams can be generated by the nodes that host the new data streams and the existing codes will remain the same. But adding new descriptors to $ST_{meaning}$ may imply different clustering results and, hence, code $\tau$ for existing descriptors may need to be changed. In fact, when $ST_{hash}$ grows extensively, we need to increase $d$ (the depth of $ST_{hash}$), which will also cause code $\tau$ for existing descriptors to change. This will have significant impact to the system, i.e., the routing tables of all the nodes in the network need to be re-established.

We avoid the need for major updates in IoT-DBN due to the addition of new data streams by keeping a larger code size to allow better tolerance of the growth. We do not have a good solution for $ST_{meaning}$ and will leave it for future research. In $ST_{hash}$, we can use a larger hash space, i.e., generate a longer hash code size. However, a hash space that is significantly larger than $\|AV_{a_i}^{sys}\|$ may result in bad routing table compression. In our solution, we generate a longer hash code, but only use the part that is appropriate for the current $AV_{a_i}^{sys}$ size. For example, if we have 100K descriptors, we may want to only use 19 bits for hash code $\tau$. But we can instead generate hash codes of 23 bits and use $d = 9$ to indicate that only $cd + 1 = 19$ bits are in use. When we increase $d$ to $d = 10$, 21 bits of the same code can be used, and the system can offer 4 times of the hash space without having to change the existing codes in the routing tables (23 bits implies tolerance of 16 times of growth). The only overhead in this solution is the propagation of the new $d$ value to every node in the IoT-DBN.

## V. WHEN TO SUMMARIZE

Assume that an st-parent $t^p$ has an st-children set $SC_{a_i}(t^p)$ of size four. In Section IV, we summarize when all four st-children code in $SCC_{a_i}(t^p)$ are present in a routing table $RT^D$ for the same neighbor. However, if the probability for getting such "**Full Sibling Code Set**" (*FSCS*) in each routing table is not sufficiently high to achieve the desired table compression rate, then we consider more aggressive summarization. Specifically, let $cov$ denote the "coverage threshold", $0 \leq cov \leq 1$. For a given st-children code set $SCC_{a_i}(t^p)$, if $RT^D$ only has $rc$ entries that match the codes in $SCC_{a_i}(t^p)$, we still summarize them to the corresponding st-parent code $\tau_i^p$ as long as $rc \geq cov \times \|SCC_{a_i}(t^p)\|$.

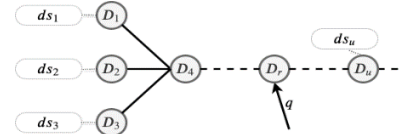

Fig. 2. Sample of network ontology

Summarizing without full FSCS can also cause misleading discovery query routing. Consider $SCC_{a_i}(t^p) = \{\tau_{i,j}, 1 \leq j \leq 4\}$. Also, consider 4 data streams $ds_1$, $ds_2$, $ds_3$, and $ds_u$, where $AN^{ds_1} = ((a_i : v^w))$, $AN^{ds_2} = ((a_i : v^x))$, $AN^{ds_3} = ((a_i : v^y))$, and $AN^{ds_u} = ((a_i : v^z))$. The code for attribute values $v^w$, $v^x$, $v^y$, and $v^z$ in $ST_x(AV_{a_i}^{sys})$ are $\tau_{i,1}$, $\tau_{i,2}$, $\tau_{i,3}$, and $\tau_{i,4}$, respectively. Consider the network topology in Fig. 2. Assume that nodes $D_1$, $D_2$, $D_3$, and $D_u$ host $ds_1$, $ds_2$, $ds_3$, and $ds_u$, respectively. $D_1$, $D_2$ and $D_3$ send the descriptors of $ds_1$, $ds_2$ and $ds_3$ to $D_4$. $RT^{D_4}$, thus, will have $(a_i:\tau_{i,1})$, $(a_i:\tau_{i,2})$ and $(a_i:\tau_{i,3})$. Assume that the coverage threshold $cov$ is set to 0.7. Following the summarization rule, $D_4$ will summarize $(a_i:\tau_{i,1})$, $(a_i:\tau_{i,2})$ and $(a_i:\tau_{i,3})$ into $(a_i:\tau^p)$, which is propagated to several intermediate nodes and reaches $D_r$. When $D_r$ receives a query $q$, where $A^q = ((a_i:v^z))$, which is converted into $(a_i:\tau_{i,4})$, misleading routing will happen i.e., $q$ may be routed to $D_4$ and get no result due to the misleading summarized descriptor $(a_i:\tau^p)$. Thus, the choice of coverage threshold $cov$ has to be done carefully. We will experimentally evaluate the impact of $cov$ to come up with a good $cov$ setting.

No matter whether we consider $cov = 1$ or a lower threshold, we need to know the size of the FSCS in order to make summarization decisions. But we do not want to store $ST_x$ on each node to provide this information. Thus, we develop methods to provide FSCS size information and they are discussed in the following subsections.

### A. FSCS Size Vector

To maintain the accurate FSCS size for summarization, we can associate the size information to each descriptor $(a_i:\tau_i)$. The only way to obtain the FSCS size is from the summarization tree by counting the siblings. To avoid needing to host $ST_x$ on each IoT-DBN node, we use the same server $STS$ discussed in Section IV to provide the size information.

**Sibling count vector (SCV)**. Let $t.ns$ be the number of siblings of node $t$ in $ST_x$. $t.ns = t^p.nc - 1$, if $t^p$ is the parent of $t$. For $ST_{meaning}$ and $ST_{hash}$, since each st-parent can only have at most $2^c$ st-children, $t.ns$ can be specified by $c$ bits. For $ST_{alph}$, if the number of all possible characters in $AV_{a_i}^{sys}$ is $N_{char}$, then $t.ns$ can be specified by $\lceil \log_2 N_{char} \rceil$ bits.

An IoT-DBN peer can query $STS$ to obtain the sibling counts for its data stream descriptors. However, if summarization has taken place and the summarized parent entry is added to the routing table, we will need to contact $STS$ to get its sibling count. To avoid such communication overhead, each node $t$ in $ST_x$ maintains a sibling count vector $t.scv$, which contains not only $t$'s sibling count, but also the sibling counts of all $t$'s ancestors in $ST_x$. Based on the requirement, $t.scv$ can be constructed by composing the $ns$ values of all its ancestors. More formally, we can construct $t.scv$ by $t.scv = (t^p.scv) \cdot (t.ns)$, where $t^p$ is the st-parent of $t$. If $t$ is the root, $t.scv = \varepsilon$ (empty string). Note that for $ST_{meaning}$, we only need to maintain $t.scv = t.ns$ for the leaf nodes because each internal node always has $2^c - 1$ siblings.

SCV is very space efficient. In $ST_{hash}$ and $ST_{meaning}$, $t.scv$ is of size $\|t.\tau_i\| - 1$. In $ST_{alph}$, it is of size $\lceil \log_2 N_{char} \rceil \times len$, where $len$ is the length of the keyword. When an IoT-DBN node advertises a descriptor $(a_i:\tau_i)$ to its neighbors, the associated SCV will be carried along with the descriptor. After summarization, the SCV of the st-parent can be derived by removing the last $y$ bits of the SCV of any st-children, where $y = c$ for $ST_{hash}$ and $y = \lceil \log_2 N_{char} \rceil$ for $ST_{alph}$. For $ST_{meaning}$, the SCV of the st-parent, after summarization, will always be $2^c - 1$.

### B. Average FSCS Size Estimation

When it is not desirable to communicate with a central server to obtain $t.scv$, we can develop a mechanism for estimating the number of siblings for each node $t$, i.e., estimate $t.ns$. Note that in $SP_{meaning}$, for each data stream descriptor, we anyway have to retrieve its code from $STS$, we can by the way retrieve its SCV. So, there is no reason to consider $t.ns$ estimation for $SP_{meaning}$. In the following, we consider $t.ns$ estimation for $ST_{hash}$ and $ST_{alph}$.

**For $ST_{hash}$**. In $ST_{hash}$, $t.ns$ depends on the choice of $c$ ($2^c$ is the bound of $t.ns$ for any $t$ in $ST_{hash}$), the number of leaf nodes in $ST_{hash}$ ($= \|AV_{a_i}^{sys}\|$), and the hash space size. Since the leaf nodes in $ST_{hash}$ are generated in the range of $[(2^c)^d, (2^c)^{d+1} - 1]$, the hash space size is $(2^c)^{d+1} - (2^c)^d$. Thus, the probability that a slot in the hash space has a leaf tree node is $\omega = \|AV_{a_i}^{sys}\|/((2^c)^{d+1} - (2^c)^d)$, assuming that the hash function in use does map a keyword space to a uniformly distributed hash code space.

Since tree nodes at different levels have different expected number of siblings, $t.ns$ also depends on $t.l$, where $t.l$ is the level $t$ is at in $ST_{hash}$. Thus, we derive $f(\omega, c, l)$, which is the expected sibling set size at level $l$ and $t.ns = f(\omega, c, t.l) - 1$ ($t.ns$ is the number of siblings of $t$, excluding $t$ itself). If $f(\omega, c, l) < 1$, then $t.ns = 0$.

Let $\gamma_l$ be the existence probability of a node in a sibling set of level $l$. Then, $2^c \times \gamma_l$ will be the expected sibling set size. We have $f(\omega, c, l) = 2^c \times \gamma_l$, and $\gamma_l = 1 - (1 - \omega)^{2^{(d-l)c}}$.

To derive $\gamma_l$, we first consider the leaf sibling set, i.e., $l = d$, where $d$ is the depth of $ST_{hash}$. Obviously, $\gamma_d = \omega$. For a parent node $t_p$ of some leaf tree nodes in $ST_{hash}$, $t_p$ does not exist if it does not have any child, which has probability $(1 - \omega)^{2^c}$. Otherwise, $t_p$ does exist. Thus, $\gamma_{d-1} = 1 - (1 - \omega)^{2^c}$. Further up the tree at level $d - 2$, a node does not exist if it does not have any child at level $d - 1$, which has probability $(1 - \omega)^{2^{2c}}$ and, hence, $\gamma_{d-2} = 1 - (1 - \omega)^{2^{2c}}$. In general, we have $\gamma_{d-l} = 1 - (1 - \omega)^{2^{lc}}$.

We round $f(\omega, c, l)$ to the nearest integer to give the estimate of the number of siblings. To evaluate the accuracy of this estimation, we get the entire English keyword ontology from the dictionary of Wordnet (including 155K keywords). From the full ontology, we randomly generated 400 sub-ontologies of different sizes. The keywords from each sub-ontology are used to construct the corresponding $ST_{hash}$ using $c = 2$. We validate the estimations against the true FSCS sizes for every node in $ST_{hash}$ and obtain the mean absolute error $MAE = 0.27$ and the mean absolute percentage error $MAPE = 9.65\%$. This gives enough accuracy for the summarization decision purpose.

**For $ST_{alph}$.** Since $ST_{alph}$ does not have parameter $c$, we express the sibling set size as $f(\omega, l)$, where $\omega = \|AV_{a_i}^{sys}\|$, and $t.ns = f(\omega, t.l) - 1$. In $ST_{alph}$, FSCS size depends on the distribution of the keywords, so it is not realistic to theoretically estimate $f(\omega, t.l)$. Thus, we use the sub-ontologies generated based on Wordnet (same sub-ontologies as discussed previously) to learn $f(\omega, l)$. We collect data points $<\omega, t.l, t.ns>$ for all $t$ in $ST_{alph}$ from 80% sub-ontologies and apply various regression learning models on them, including linear, quadratic, cubic, logarithmic, rational, and exponential. The results show that $f(\omega, l)$ has almost no dependency on $\omega$. So, we derive $f(l)$ and the model with the lowest error rate is $f(l) = 26 \times l^{-2}$.

Similar to $ST_{hash}$, $f(l)$ is rounded to the nearest integer to get the estimation. We apply the estimation model to the 20% remaining sub-ontologies to validate the estimation accuracy. Error analysis shows $MAE = 0.22$ and $MAPE = 5.79\%$.

**Remarks**. To enable FSCS size estimation under $SP_{hash}$, each node in IoT-DBN needs to know the values of $c$, $d$, $\omega$ and $l$. We assume that the system is relatively stable. Thus, $c$ and $d$ can be predetermined for the network. $\omega$ may change dynamically and we assume that a growth function for the number of data streams ($\|AV_{a_i}^{sys}\|$) is given. If the system needs to update $d$ due to data growth, a leader may initiate a flooding message to inform each IoT-DBN node to change $d$ and the corresponding hash code length. The descriptor for a real data stream always has $l = d$. For FSCS size estimation for $SP_{alph}$, we only need to know $l$, where $l = length(\tau_i)$. For $SP_{hash}$, $l$ can be derived from code $\tau_i$.

## VI. EXPERIMENTAL STUDY

We simulated an IoT network using the technique of Fast Network Simulation Setup (FNSS) toolchain [18]. Networks of different sizes (#nodes), from 1K to 30K nodes, are generated. Each node in the network can have 2 to 10 neighbors (following a weighted uniform distribution). We crawled the web and obtained about 100K IoT data stream annotations that include 15 attributes and 55K unique keywords (879K keywords in total). The raw data can be downloaded from [7].

Data descriptors for each object, with some random replications, are uniformly randomly distributed to the nodes near and at the edge of the network. A query is generated by uniformly randomly selecting a data stream and uniformly randomly selecting some of its descriptors to include. For each query, a uniformly randomly selected node in the network is designated to initiate it.

We implemented AdvP and QFP to study the effectiveness of various summarization algorithms in reducing routing table size and their impact to discovery query routing performance. We also compare the performance of representative algorithms in major p2p data discovery approaches, including (1) our solutions and (2) the GSD-like peer-to-peer algorithm. Two simple DHT-based approaches are also considered, including (3) the psKey technique [1] and (4) the Mkey algorithm [12]. The case of (2) represents a set of unstructured routing algorithms that do not summarize for keywords, including GSD [4] and CBCB [5]. For fairness, the information cache for GSD are already built up before discovery queries are issued. (3) and (4) have been discussed in Section II. For (3), we derive all the subsets of the set of keyword-based descriptors for each data stream and hash them to the DHT. In (4), the Mkey structure, which is a hybrid of DHT and unstructured cluster, is simulated. Bloom filters of size 128 bits is used to code the descriptors of each data stream and is split into hash codes for the DHT for storing the data stream indices.

### A. Configuring Summarization Algorithms

We study the impact of the parameter settings in various algorithms, including (a) the coverage threshold $cov$, (b) the degree bound for the summarization trees, $2^c$. The impact is measured in terms of (i) RT-size (the number of routing table entries); (ii) Traffic (the average number of messages, i.e., the total number of hops); and (iii) Latency (the number of hops for each query to successfully get its responses from all the routing directions). The hop bound for advertisement $b^{ad}$ and for query forwarding $b^q$ are set to $\infty$.

**(a) coverage threshold**. Coverage plays an important role for summarization. We evaluate the impact on RT-size, Traffic, and Latency due to different coverage thresholds $cov$ in various summarization policies considering different network sizes. Other parameters are set to $2^c = 4$ and $\alpha = 1.0$. Due to space limit and since the results for different summarization policies are similar, we only present the results for $SP_{meaning}$ in Fig. 3.

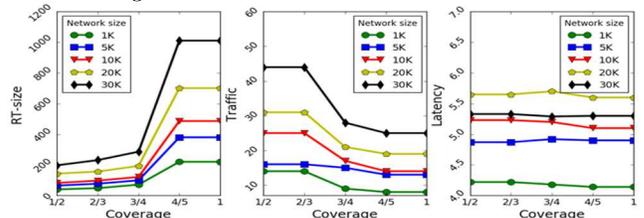

Fig. 3. Coverage Impact for $SP_{meaning}$

Clearly, $cov = 3/4$ gives the best tradeoff between RT-size and Traffic for all policies. When $cov \geq 3/4$, RT-sizes are very reasonable, but Traffic becomes significantly higher. When $cov < 3/4$, Traffic is good, but RT-size rises sharply. Although $cov = 3/4$ yields the best tradeoff, we use $cov=1$ in experiments to show the effectiveness of summarization.

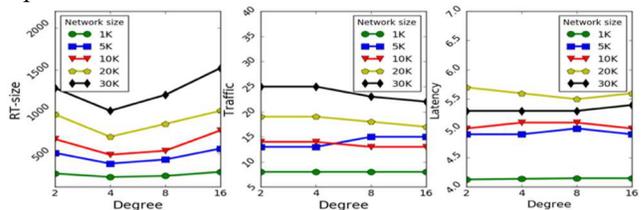

Fig. 4. Degree comparison for $SP_{meaning}$

**(b) Degree bound for the summarization trees**. We study the impact of different settings of $2^c$ and execute AdvP and QFP with different summarization policies and different network sizes. The other parameters are set as $cov = 1.0$ and

$\alpha = 1.0$. Same reason of space limitation as above, we only present the results for $SP_{meaning}$ in Fig. 4.

As can be seen, when $2^c = 4$, routing table has the most effective compression. RT-size rises when $2^c > 4$ because the probability of having a sufficient number of siblings for summarization becomes lower. For $2^c = 2$, we further investigated the reason for the higher RT-size and found that summarization happens more frequently at the leaf nodes, but less frequently at the higher layers, resulting in a lower overall compression rates compared to the case of $2^c = 4$. Traffic for increasing $c$ drops slowly with increasing $c$ and Latency only differs marginally with different $c$ settings. Thus, we choose $2^c = 4$ for all summarization policies.

### B. Misleading Routing

In addition to studying the overall Traffic and Latency, we also investigate the misleading routing problem due to (a) the coverage threshold in summarization and (b) the mechanisms for computing coverage. Understanding misled traffic can help gain fine-grained insights on the impacts of the design decisions in discovery routing methods. The parameters that are fixed in this section include $2^c = 4, \alpha = 1.0$.

**(a) Coverage threshold**. The percentages of misled messages in overall Traffic due to $cov \neq 1$ are shown in Fig. 5. When $cov = 1.0$, there is no misled messages due to coverage. The 2% misled messages are due to indexing by individual descriptors without identifying their data streams. For example, consider the descriptors of 3 data streams: $AN^{ds_1} = ((a_1: v^w), (a_2: v^x))$, $AN^{ds_2} = ((a_1: v^y), (a_2: v^z))$, and $AN^{ds_3} = ((a_1: v^w), (a_2: v^z))$. A node $D$ advertising descriptors in $AN^{ds_1}$ and $AN^{ds_2}$ may cause misled routing toward $D$ for queries looking for $AN^{ds_3}$. When decreasing coverage threshold $cov$, percentage of misleading message increases. $cov = 3/4$ yeilds the best tradeoff, conforming to the Traffic results in Fig. 3.

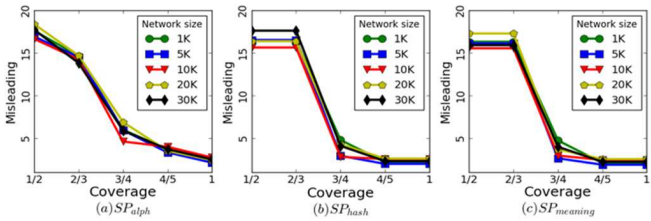

Fig. 5. Misleading Impact for policies

**(b) SCV vector vs FSCS size estimation**. Maintaining SCV in the routing table can incur communication costs for SCV retrieval, but it allows an accurate coverage computation. We analyzed the accuracy of FSCS size estimation and study the impact of this inaccuracy toward the percentage of misled messages and the results are shown in Fig. 6.

Since we set $cov = 1.0$, using SCV will not cause misled routing and the 2% misled messages comes from indexing by individual descriptors without identifying their data streams. Let $E$ and $A$ denote the estimated and the actual FSCS sizes. $E > A$ will not yield misleading messages. If $E < A$, coverage may be underestimated, and may result in some misleading routing. We analyzed the results and observed that 0.3% additional misleading messages are due to $E < A$, which subsequently causes the compression ratio to rise from 12.9 times to 13.5 times.

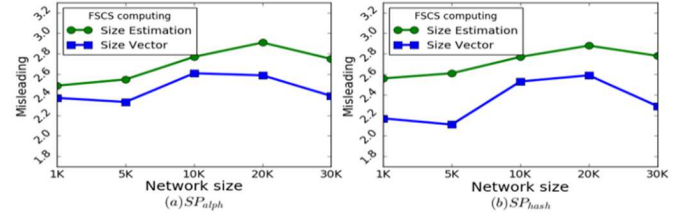

Fig. 6. Misleading impact for FSCS computing

### C. Summarization Effectiveness

We now study the effectiveness of our three summarization algorithms. We use the same random object distribution (Random) discussed in the beginning of this section as well as the regionalized object distribution (Region) discussed in Section IV to evaluate the algorithms. For Region distribution, we automatically divide the network into a number of "regions" based on the traversal order. We also cluster the data streams based on the similarity of their descriptors. Then, we distribute objects in the same cluster to the same region to create the regionalized keyword distribution effect. In the experiment, we use 7 regions and 14 clusters. We use the no summarization case (called $nSum$) as the baseline for evaluation. We set $cov = 1.0$. Fig. 7 shows the results.

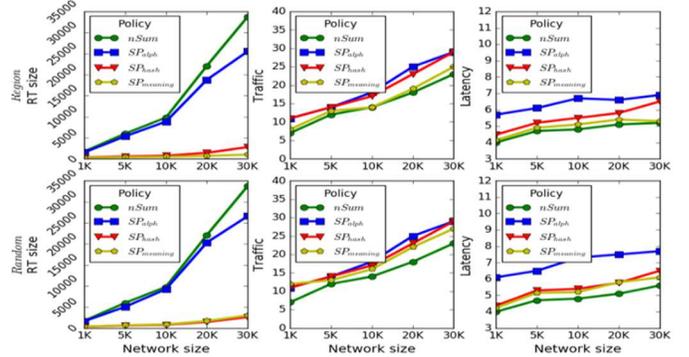

Fig. 7. Policy comparison

In a relatively large scale network (30K node), $ST_{meaning}$ and $ST_{hash}$ achieves 11.1 and 12.9 compression ratios, respectively, in Random distribution, and 33.3 and 12.1 compression ratios, respectively, for Region distributions. The total Traffic increases from $nSum$ due to summarization by $SP_{meaning}$ are 7% in Region and 10% in Random distribution, by $SP_{hash}$ is 12% (only considered Random), and by $SP_{alph}$ is 15%. The increases in Traffic is acceptable considering the at least 11 times reduction in RT-size. In terms of Latency, $SP_{hash}$ and $SP_{meaning}$ have roughly a 10% increase from $nSum$ while the increase in Latency by $SP_{alph}$ is 30%.

Note, the comparison is based on unlimited routing table size. If we confine the RT size, some routing information may be forced to be thrown away in existing approaches, which will significantly increase their Traffic and Latency results.

## D. Compare Data Discovery Approaches

We further compared the performance of five data discovery approaches discussed in the beginning of this section. We choose $SP_{hash}$ to represent our summarization techniques. For the GSD-like approach, we consider unbounded cache size (GSD-like-nBound) and bounded cache size (GSD-like-Bounded). In the latter case, GSD cache is bounded by the same size as the RT-size in $SP_{hash}$. The bound is achieved by using LRU to eject the extra entries. The results for Traffic and Latency are shown in Fig. 8.

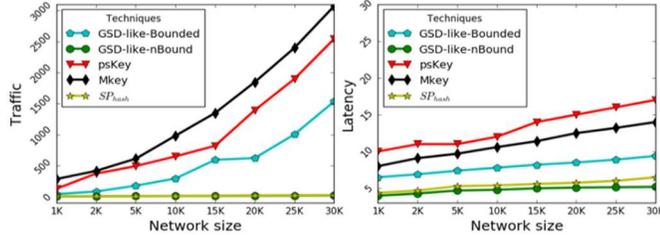

Fig. 8. Comparison of major data discovery approaches

As can be seen, our $SP_{hash}$ algorithm outperforms all other data discovery approaches except for GSD-like-nBound. DHT based approaches (psKey and Mkey) require routing to find the node hosting the location information and they also have wasted communications with nodes that may not have the queried results and Chord routing has inherited extra hops for each intermediate destination. Thus, they impose a much higher overhead in Traffic and Latency. In GSD-like-Bounded, useful routing information are being thrown away instead of summarized and, hence, it has worse Traffic and Latency than $SP_{hash}$. Though GSD-like-nBound solution has a slightly better Traffic and Latency than $SP_{hash}$, its cache size will be 23.5 times larger than that of $SP_{hash}$.

With increasing network size, all other algorithms besides the p2p ones, especially the DHT based approaches, have drastically increasing Traffic and Latency. At a very large scale IoT-DBN, (for instance, 30K nodes) both the p2p approaches have 2 to 4 times lower Latency and 50 to 100 times lower Traffic in comparison with psKey and Mkey approaches. GSD-like-Bounded having the same RT-size with our $SP_{hash}$ yields 20 times higher Traffic and 2 times higher Latency.

## VII. CONCLUSION

We explored the p2p data discovery problem in very large and growing IoT networks. Specifically, we delved deeply into the summarization techniques for MAA (and multi-keyword) based IoT data discovery. To achieve space efficiency for discovery routing, we not only analyzed potential summarization strategies, but also designed the novel coding scheme for each summarization strategy and explored the coverage issues. Our work is the first to make summarization for keyword based discovery routing truly feasible. For experimental studies, we crawled the web to collect a large dataset (posted on GitHub). Experimental results demonstrate the effectiveness of our summarization solutions and show that they outperform other p2p discovery solutions significantly.